\documentclass{article}
\usepackage{graphicx}

\newcommand{\sfrac}[2]{\mbox{\footnotesize $\frac{#1}{#2}$}}

\begin{document}

\begin{center}
{\large DSEs, THE PION, AND RELATED MATTERS}\\
\medskip
ANDREAS KRASSNIGG$^{1}$ AND CRAIG D.\ ROBERTS$^{\,1,2}$\\
\smallskip
$^1${\em Physics Division, Argonne National Laboratory, Argonne IL 60439,
USA}\\
$^2${\em Fachbereich Physik, Universit\"at Rostock, D-18051 Rostock, Germany}
\end{center}

\hspace*{-\parindent}  We recapitulate on aspects of Dyson-Schwinger equation
studies relevant to pseudoscalar mesons: lattice confirmation of the DSE
prediction that propagators are nonperturbatively dressed in the infrared; and
exact results, e.g., in the chiral limit the leptonic decay constant vanishes
for every pseudoscalar meson except the pion.

\bigskip

\hspace*{-\parindent}{\it PACS numbers}: 12.38.Aw, 14.40.Aq, 14.65.-q, 24.85.+p
\vspace*{0.1em}

\hspace*{-\parindent}{\it Keywords}: Dynamical chiral symmetry breaking,
Dyson-Schwinger Equations, Pion properties.
\medskip

\begin{center} {\large\it 1.~Gap equation}
\end{center}
The vast body of pion data available provides compelling evidence that this
composite particle is a Goldstone mode of the strong interaction associated
with the dynamical breaking of chiral symmetry.  Therefore a legitimate
understanding of pion observables, including its mass, decay constants and form
factors, requires that an approach possess a well-defined and valid chiral
limit.  This is impossible without a detailed grasp of the connection between
current- and constituent-quarks.

In QCD the running quark mass is obtained from the solution of
\begin{equation}
\label{gendse} S^{-1}(p) = Z_2 \,(i\gamma\cdot p + m_{\rm bare}) +\, Z_1
\int^\Lambda_q \, g^2 D_{\mu\nu}(p-q) \frac{\lambda^a}{2}\gamma_\mu\, S(q)\,
\Gamma^a_\nu(q;p) \,.
\end{equation}
This is the Dyson-Schwinger equation (DSE) for the dressed-quark self energy
or, equivalently, QCD's \textit{gap equation}, and it is a keystone in
understanding dynamical chiral symmetry breaking (DCSB) and the relation
between current- and constituent-quarks.  On the right hand side of Eq.\
(\ref{gendse}): $D_{\mu\nu}(k)$ is the dressed-gluon propagator;
$\Gamma^a_\nu(q;p)$ is the dressed-quark-gluon vertex; $m_{\rm bare}$ is the
$\Lambda$-de\-pen\-dent current-quark bare mass; and $\int^\Lambda_q :=
\int^\Lambda d^4 q/(2\pi)^4$ represents a translationally-invariant
regularisation of the integral, with $\Lambda$ the regularisation mass-scale.
In addition, $Z_{1,2}(\zeta^2,\Lambda^2)$ are the quark-gluon-vertex and quark
wave function renormalisation constants, which depend on $\Lambda$ and the
renormalisation point, $\zeta$, as does the mass renormalisation constant $
Z_m(\zeta^2,\Lambda^2) = Z_4(\zeta^2,\Lambda^2)/Z_2(\zeta^2,\Lambda^2) $. The
solution of Eq.\ (\ref{gendse}) has the form
\begin{equation}
 S^{-1} (p) =  i \gamma\cdot p \, A(p^2,\zeta^2) + B(p^2,\zeta^2) \\
 \equiv \frac{1}{Z(p^2,\zeta^2)}\left[ i\gamma\cdot p + M(p^2)\right],
\label{sinvp}
\end{equation}
where $M(\zeta^2) \equiv m(\zeta):= m_{\rm
bare}(\Lambda)/Z_m(\zeta^2,\Lambda^2)$ is the running quark mass.

\begin{figure}[t]
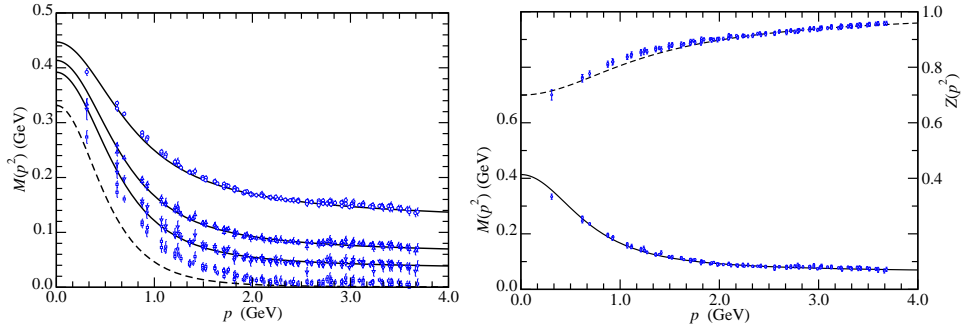

\leftline{\includegraphics[height=12.1em]{fig1.eps}} \vspace*{-12.1em}
\rightline{\includegraphics[height=12.1em]{fig2.eps}}
\caption{\label{mandarcflattice} \textit{Left Panel} -- Dashed-curve: gap
equation's solution in the chiral limit; solid curves: solutions for $M(p^2)$
obtained using the current-quark masses in Eq.\ (\protect\ref{amvalues}). (From
Ref.\ [\protect\ref{mandar}].)  Data, upper three sets: lattice results for
$M(p^2)$ in GeV at $am$ values in Eq.\ (\protect\ref{amvalues}); lower points
(boxes): linear extrapolation of these results [\protect\ref{bowman2}] to $a m
=0$. \textit{Right Panel} -- Dashed curve, $Z(p^2)$, and solid curve, $M(p^2)$
calculated from the gap equation with $m(\zeta)=55\,$MeV
[\protect\ref{mandar}].  Data, quenched lattice-QCD results for $M(p^2)$ and
$Z(p^2)$ obtained with $am = 0.036$ [\protect\ref{bowman2}].  ($Z(p^2)$ is
dimensionless.)}
\end{figure}

The behaviour of the nonperturbative solution of QCD's gap equation is a
longstanding prediction of DSE studies
[\ref{revbasti},\ref{revreinhard},\ref{revpieter}], and typical results are
illustrated in Fig.~\ref{mandarcflattice}.  One critical feature is that so
long as the kernel of the gap equation has sufficient integrated strength at
infrared momenta, one obtains a nonzero running quark mass even in the chiral
limit:
\begin{equation}
\label{chirallimit} Z_{4}(\zeta,\Lambda) \, m(\zeta) \equiv 0\,,\;  \Lambda \gg
\zeta \,.
\end{equation}
This effect is DCSB.  It is impossible at any finite order of perturbation
theory and apparent in Fig.\ \ref{mandarcflattice}.

The dressed-quark propagator can be calculated in lattice-regularised QCD.
Results are available in the quenched truncation, and depicted in Fig.\
\ref{mandarcflattice} are those of Ref.\ [\ref{bowman2}] obtained with the
current-quark masses ($\zeta = 19\,$GeV)
\begin{equation}
\label{amvalues}
\begin{array}{l|lll}
a\,m_{\rm lattice} & 0.018 & 0.036 & 0.072 \\\hline m(\zeta) ({\rm GeV}) &
0.030 & 0.055 & 0.110
\end{array}\,.
\end{equation}
The precise agreement with DSE results is not accidental.  The essential
agreement between lattice results and DSE predictions was highlighted in Refs.\
[\ref{pctlattice}] but Ref.\ [\ref{mandar}] pursued a different goal.  Only
recently has reliable information about the gap equation's kernel at  infrared
momenta begun to emerge, in the continuum [\ref{alkofer}] and on the lattice
[\ref{latticegluon}]. Reference [\ref{mandar}] therefore employed an
\textit{Ansatz} for the infrared behaviour of the gap equation's kernel in
order to demonstrate that it is possible to correlate lattice results for the
gluon and quark Schwinger functions via QCD's gap equation.   This required the
gap equation's kernel to exhibit infrared enhancement over and above that
observed in the gluon propagator alone, which could be attributed to an
amplification of the dressed-quark-gluon vertex whose magnitude is consistent
with that observed in quenched lattice estimates of this three-point function
[\ref{ayse}].
\medskip

\begin{center}
{\large\it 2.~Hadrons}
\end{center}
It is evident that reliable knowledge of QCD's two-point functions (the
propagators for QCD's elementary excitations) is available.  Direct comparison
with experiment requires an equally good understanding of bound states.
Progress here has required the evolution of an understanding of the intimate
connection between symmetries and DSE truncation schemes.  The best known
scheme is the weak coupling expansion, which reproduces every diagram in
perturbation theory.  This scheme is valuable in the analysis of large momentum
transfer phenomena because QCD is asymptotically free. However, it precludes
any possibility of obtaining nonperturbative information, and bound state
phenomena are intrinsically nonperturbative.

The properties of the pion are profoundly connected with DCSB, and chiral
symmetry and its breaking are expressed through the axial-vector Ward-Takahashi
identity ($k_\pm = k \pm P/2$, $\{\tau^j,j=1,2,3\}$ are the Pauli matrices)
\begin{equation}
\label{avwtim} P_\mu \Gamma_{5\mu}^j(k;P)  = {\cal S}^{-1}(k_+) i
\gamma_5\frac{\tau^j}{2} +  i \gamma_5\frac{\tau^j}{2} {\cal S}^{-1}(k_-) - i
{\cal M}(\zeta) \,\Gamma_5^j(k;P) - \Gamma_5^j(k;P)\,i {\cal M}(\zeta).
\end{equation}
This identity connects the axial-vector vertex: $\Gamma_{5\mu}^j(k;P)$, $P$ is
the total momentum, with the dressed quark propagator: ${\cal S} = {\rm
diag}[S_u,S_d]$, the pseudoscalar vertex: $\Gamma_{5}^j(k;P)$, and the
current-quark mass matrix: ${\cal M}(\zeta) = {\rm
diag}[m_u(\zeta),m_d(\zeta)]$.  The propagator satisfies the gap equation but
the vertices are determined by inhomogeneous Bethe-Salpeter equations; e.g.,
\begin{eqnarray}
\label{genave} \left[\Gamma_{5\mu}^j(k;P)\right]_{tu} &= &Z_2 \,
\left[\gamma_5\gamma_\mu \frac{\tau^j}{2}\right]_{tu} \,+ \int^\Lambda_q \,
[\chi_{5\mu}^j(q;P)]_{sr} \,K^{rs}_{tu}(q,k;P)\,,
%
\end{eqnarray}
wherein $\chi^j_{5\mu}(q;P):= {\cal S}(q_+) \Gamma^j_{5\mu}(q;P) {\cal S}(q_-)$
and $K(q,k;P)$ is the dressed-quark-antiquark scattering kernel.  The
importance of DCSB entails that any truncation useful in understanding low
energy phenomena must be nonperturbative and preserve Eq.\ (\ref{avwtim}),
without fine tuning. This nontrivial constraint cannot be satisfied without an
intimate connection between $K(q,k;P)$ and the gap equation's kernel.

One systematic truncation scheme has been identified that explicates this
connection and hence preserves QCD's global symmetries [\ref{detmold}]. It is a
dressed-loop expansion of the dressed-quark-gluon vertices that appear in the
half-amputated dressed-quark-anti\-quark scattering matrix: $S^2 K$.  The
leading order term is the renormalisation-group-improved rainbow-ladder
truncation, which underlies a one-parameter model of the quark-quark
interaction used successfully in \textit{ab initio} calculations of vector and
flavour nonsinglet pseudoscalar meson properties [\ref{revpieter}].

%
The existence of a nonperturbative, systematic and symmetry preserving
truncation scheme enables exact results to be proved.  For example, it is a
general feature of QCD that the axial-vector and pseudoscalar vertices exhibit
poles whenever $P^2= - m^2_{\pi_n}$, where $m_{\pi_n}$ is the mass of the pion
or any of its radial excitations.\footnote{The hadron spectrum exhibits a
sequence of $J^{PC}=0^{-+}$ mesons, with $\pi(140)$ being the lowest mass
entry. In quantum mechanical models the other members of this sequence:
$\pi(1300)$, $\pi(1800)$, \ldots\,, are described as radial excitations of the
$\pi(140)$. Aspects of this interpretation persist in Poincar\'e covariant
studies in quantum field theory and hence we retain the nomenclature.} This can
be expressed for the axial-vector vertex via
\begin{eqnarray}
\nonumber \left. \Gamma_{5 \mu}^j(k;P)\right|_{P^2+m_{\pi_n}^2 \approx 0}  &=&
\frac{\tau^j}{2} \gamma_5 \left[ \gamma_\mu F_R(k;P) + \gamma\cdot k k_\mu
G_R(k;P) - \sigma_{\mu\nu} \,k_\nu\, H_R(k;P)
\right]\\
& & + \, \tilde\Gamma_{5\mu}^{j}(k;P) + \frac{f_{\pi_n} \, P_\mu}{P^2 +
m_{\pi_n}^2} \Gamma_{\pi_n}^j(k;P)\,, \label{genavv}
\end{eqnarray}
where: $F_R$, $G_R$, $H_R$ and $\tilde\Gamma_{5\mu}^{i}$ are regular as $P^2\to
-m_{\pi_n}^2$, $P_\mu \tilde\Gamma_{5\mu}^{i}(k;P) \sim {\rm O }(P^2)$ and
nonsingular; $\Gamma_{\pi_n}^j(k;P)$ is the $0^{-+}$ bound state's
Bethe-Salpeter amplitude:
\begin{eqnarray}
\nonumber \Gamma_{\pi_n}^j(k;P)& =&  \tau^j \gamma_5 \left[ i E_{\pi_n}(k;P)
+ \gamma\cdot P F_{\pi_n}(k;P) \right. \\
&&  \left.  +\,  \gamma\cdot k \,k \cdot P\, G_{\pi_n}(k;P) +
\sigma_{\mu\nu}\,k_\mu P_\nu \,H_{\pi_n}(k;P)  \right], \label{genpibsa}
\end{eqnarray}
which is determined by the homogeneous Bethe-Salpeter equation
\begin{eqnarray}
\label{genbsepi} \left[\Gamma_{\pi_n}^j(k;P)\right]_{tu} &=&  \int^\Lambda_q
\,[\chi_{\pi_n}^j(q;P)]_{sr} \,K^{rs}_{tu}(q,k;P)\,;
\end{eqnarray}
and $f_{\pi_n}$ is the pseudoscalar meson's leptonic decay constant
\begin{equation}
\label{fpin} f_{\pi_n} \,\delta^{ij} \,  P_\mu = Z_2\,{\rm tr} \int^\Lambda_q
\sfrac{1}{2} \tau^i \gamma_5\gamma_\mu \chi_{\pi_n}^j(q;P)\,,
\end{equation}
where the trace is over colour, flavour and spinor indices. Equation
(\ref{fpin}) is the expression in quantum field theory for the
pseudo\textit{vector} projection of the meson's Bethe-Salpeter wave function
onto the origin in configuration space.

The analogous expression in the case of the pseudoscalar vertex is
\begin{eqnarray}
 \left. i \Gamma_{5}^j(k;P)\right|_{P^2+m_{\pi_n}^2 \approx 0} &= &
\mbox{regular\ terms} + \frac{ \rho_{\pi_n} }{P^2 + m_{\pi_n}^2}\,
\Gamma_{\pi_n}^j(k;P)\,,\label{genpvv}\\
\label{cpres} i  \rho_{\pi_n}\!(\zeta)\, \delta^{ij} & = &Z_4\,{\rm tr}
\int^\Lambda_q \sfrac{1}{2} \tau^i \gamma_5 \chi_{\pi_n}^j(q;P)\,.
\end{eqnarray}
Equation (\ref{cpres}) expresses the pseudo\textit{scalar} projection of the
meson's Bethe-Salpeter wave function onto the origin in configuration space.

Inserting Eqs.\ (\ref{genavv}), (\ref{genpvv}) into Eq.\ (\ref{avwtim}), and
subsequently equating pole terms, one obtains the model-independent result
[\ref{mrt98}]
\begin{equation}
\label{gmorgen} f_{\pi_n} m_{\pi_n}^2 = [ m_u(\zeta) + m_d(\zeta) ] \,
\rho_{\pi_n}(\zeta)\,.
\end{equation}

In the chiral limit, Eq.\ (\ref{chirallimit}), the axial-vector Ward-Takahashi
identity becomes
\begin{equation}
\label{avwti0} P_\mu \Gamma_{5\mu}^j(k;P)  = {\cal S}^{-1}(k_+) i
\gamma_5\frac{\tau^j}{2} +  i \gamma_5\frac{\tau^j}{2} \,{\cal S}^{-1}(k_-) .
\end{equation}
Assume that chiral symmetry is dynamically broken so that the dressed-quark
propagator has a nonzero Dirac-scalar term; viz., $B\neq 0$ in Eq.\
(\ref{sinvp}).  It then follows [\ref{mrt98}] that there is a massless
pseudoscalar bound state, $m_{\pi_0}=0$, for which
\begin{equation}
\label{gtrelation} f_{\pi_0}^0 E_{\pi_0}(k;0) = B(k^2)\,,
\end{equation}
with similar relations for the other scalar functions in Eq.\ (\ref{genpibsa}).
This is clearly the ground state and finiteness of the right hand side in Eq.\
(\ref{gtrelation}) entails $f_{\pi_0}^0\neq 0$. Moreover, for the ground state
pion in the chiral limit [\ref{mrt98}]
\begin{equation}
\rho_{\pi_0}(\zeta) \to \rho_{\pi_0}^0(\zeta)= \frac{1}{f_{\pi_0}^0}\,
Z_4\,{\rm tr} \int^\Lambda_q S^0(q) = - \frac{1}{f_{\pi_0}^0}\,\langle \bar q q
\rangle_\zeta^0\,,
\end{equation}
where $\langle \bar q q \rangle^0$ is the vacuum quark condensate.  Hence the
Gell-Mann--Oakes--Renner relation for the ground state pion appears as a
corollary of Eq.\ (\ref{gmorgen}).
Another important corollary of Eq.\ (\ref{gmorgen}), valid for pseudoscalar
mesons containing at least one heavy-quark, is described in Ref.\
[\ref{hqlimit}].

It is plain that  $m_{\pi_{n\neq 0}}> m_{\pi_{n=0}}$ for all radial excitations
and hence in the chiral limit $m_{\pi_{n\neq 0}}>0$.  In this limit it is
impossible to avoid the fact that the absence of a pole contribution on the
right hand side of Eq.\ (\ref{avwti0}) forces
\begin{equation}
f^0_{\pi_{n\neq 0}} = 0 \,;
\end{equation}
viz., in the chiral limit the leptonic decay constant vanishes for every one of
the pion's radial excitations.\footnote{A discussion of this result in chiral
quark models is presented in Ref.\ [\protect\ref{volkov}].  We thank M.K.\
Volkov and V.L.\ Yudichev for bringing this to our attention.}   In general
\begin{equation}
\label{fpinfpi0} \frac{f_{\pi_{n}}}{f_{\pi_{0}}} =
\frac{m_{\pi_{0}}^2}{m^2_{\pi_{n}}} \, \frac{\rho_{\pi_n}}{\rho_{\pi_0}}\,.
\end{equation}
\medskip

\begin{center}
{\large\it 3.~Quantitative illustration}
\end{center}
The manner in which these results are realised in QCD can be illustrated using
the one-parameter renormalisation-group-improved ladder model for the
quark-antiquark scattering kernel introduced in Ref.\ [\ref{maristandy1}], and
reviewed in Ref.\ [\ref{revpieter}].  One first solves the gap equation, Eq.\
(\ref{gendse}), whose solution is required to complete the specification of the
Bethe-Salpeter equation, Eq.\ (\ref{genbsepi}), and then solves this for the
pion and and its first radial excitation.

The homogeneous Bethe-Salpeter equation is an eigenvalue problem, with the
bound state masses: $P^2=-m_{\pi_n}^2$, being the eigenvalues.  We want only
the first two eigenvalues and eigenvectors.  They can be obtained from the
modified equation
\begin{equation}
\label{genbsepilambda} l(P^2)\, \left[\Gamma_{\pi_n}^j(k;P)\right]_{tu} =
\int^\Lambda_q \,[\chi_{\pi_n}^j(q;P)]_{sr} \,K^{rs}_{tu}(q,k;P)\,,
\end{equation}
with $l(P^2)$ a scalar, which has a solution for every value of $P^2$ and can
therefore be solved by iteration. To explain this, consider the equation
written in the form
\begin{equation}
l (P^2) \, |g\rangle = M(P^2)\, |g\rangle \,,
\end{equation}
where the matrix $M(P^2)$ denotes the full kernel of the Bethe-Salpeter
equation. One fixes a value of $P^2$ and ``guesses'' a solution: $ |g(0)\rangle
$. The kernel, $M(P^2)$, has a complete set of real eigenvectors $|g_i\rangle$
with eigenvalues $\lambda_i$, ordered such that $\lambda_0 > \lambda_1
> \ldots$, and therefore
\begin{equation}
|g(0)\rangle = \sum_{i=0}^\infty\, a_i |g_i\rangle\,,
\end{equation}
where $a_i$ are real constants and the vector is canonically normalised. It is
clear that
\begin{equation}
M(P^2)^N\,|g(0)\rangle = \sum_{i=0}^\infty\, \lambda_i^N\, a_i |g_i\rangle\ =
\lambda_0^N \, \left\{ a_0 |g_0\rangle\ + \sum_{i=1}^\infty\,
\frac{\lambda_i^N}{\lambda_0^N} \, a_i |g_i\rangle\right\}
\end{equation}
and hence for sufficiently large $N$,
\begin{equation}
M(P^2)^{N+1}\,|g(0)\rangle \approx \lambda_0  \, M(P^2)^{N}\,|g(0)\rangle .
\end{equation}
Thus repeated operation of the kernel on the initial ``guess''  produces the
largest eigenvalue, $l_0 (P^2)=\lambda_0$, and its associated eigenvector to
any required accuracy.

One completes this exercise for a range of values of $P^2$ and thereby obtains
a trajectory $l_0(P^2)$ that maps the $P^2$-evolution of the integral
equation's largest eigenvalue. It is then straightforward to find that $P^2$
for which $l_0(P^2)=1$.  This is the solution of Eq.\ (\ref{genbsepi}) so that
$P^2= -m_{\pi_0}^2$ and the associated eigenvector is the ground state pion's
Bethe-Salpeter amplitude.

The procedure can also be applied to determine the first excited state. One
fixes $P^2 < -m_{\pi_0}^2$, and finds the largest eigenvalue and associated
eigenvector for this new mass-scale.   That completed, one again ``guesses''
the Bethe-Salpeter amplitude but now projects out the eigenvector associated
with the largest eigenvalue at this $P^2$:
\begin{equation}
\label{project} |\tilde g(0)\rangle =  |g(0) \rangle- | g_0 \rangle
\,\frac{1}{\langle g_0 | g_0\rangle} \, \langle g_0 |g(0)\rangle\,.
\end{equation}
The iterative procedure then applied as before to $|\tilde g(0)\rangle$ yields
the second largest eigenvalue and its associated eigenvector. One thereby
obtains the $P^2$-evolution of the second largest eigenvalue, $l_1(P^2)$.  The
solution of $l_1(P^2)=1$ gives the mass of the first excited state and at this
$P^2$ the eigenvector is that state's Bethe-Salpeter amplitude.  Any finite
number of excited states can be studied in this way.

We have not yet specified the meaning of the inner product used implicitly in
Eq.\ (\ref{project}).  For models in the class characterised by the
rainbow-ladder truncation
\begin{equation}
\langle h| g \rangle :=  {\rm tr} \int_q^\Lambda\, h(q;-P) \,S(q_+)\, g(q;P)\,
S(q_-)\,.
\end{equation}
The condition $\langle h | g \rangle = 0$ then expresses a statement that the
momentum-dependent $g\to h$ vacuum polarisation (overlap amplitude) vanishes at
$P^2$, which is akin to saying that the $P^2$-dependent mass-mixing matrix for
the states $g$, $h$ does not possess off-diagonal terms.

We have obtained the mass and amplitude for the ground state pion, using the
complete expression in Eq.\ (\ref{genpibsa}), and found: $ f_{\pi_0} =
0.092$\,{\rm GeV};  $m_{\pi_0} = 0.14$\,{\rm GeV}; $\rho_{\pi_0} = (0.81\,{\rm
GeV})^2$, at a current-quark mass $m_d(1\,{\rm GeV})= m_u(1\,{\rm GeV})=
5.5\,$MeV, reproducing the results in Ref.\ [\ref{maristandy1}].

Our study of the first excited state is in its early stages and hitherto we
have only employed the leading amplitude, $E_{\pi_1}$, in Eq.\
(\ref{genpibsa}), and therewith obtained an estimate for the mass:
\begin{equation}
m_{\pi_{n=1}} \approx 1.1\,{\rm GeV}.
\end{equation}
Assuming $\rho_{\pi_{n=1}} \leq \rho_{\pi_0}$, which is supported by our
preliminary estimates, it follows from Eq.\ (\ref{fpinfpi0}) that at the
physical current-quark mass
\begin{equation}
f_{\pi_{n=1}} \leq 0.016\,f_{\pi_{n=0}} = 1.5\,{\rm MeV}.
\end{equation}
\medskip

\begin{center}
{\large\it 4.~Epilogue}
\end{center}
We have necessarily been brief.  There are many other applications of interest
to this community, among them the \textit{ab initio} calculation of
electromagnetic and transition pion form factors, and a calculation of the
pion's valence-quark distribution function whose discrepancy with extant data
raises difficult questions. These and other studies are reviewed in Ref.\
[\ref{revpieter}].
A pressing contemporary challenge is the extension of the framework to the
calculation of baryon observables, aspects of which are beginning to be
understood [\ref{piN}].

\medskip

\begin{center}
\textit{Acknowledgments}
\end{center}
CDR thanks the organisers for their assistance and, in particular, Dubravko
Kla\-bu\-\v{c}ar for his kindness and hospitality.  We acknowledge useful
conversations with D.B.\ Blaschke, P.C.\ Tandy, M.K.\ Volkov and V.L.\
Yudichev.
This work was supported by: the Austrian Research Foundation FWF,
Erwin-Schr\"odinger-Stipendium no.\ J2233-N08; the Department of Energy,
Nuclear Physics Division, contract no.\ W-31-109-ENG-38; and the A.v.\
Humboldt-Stiftung via a F.W.\ Bessel Forschungspreis.
%
\medskip

\begin{center}
References
\end{center}
\begin{enumerate}
\item \label{revbasti} C.\,D.\ Roberts and S.\,M.\ Schmidt, Prog.\ Part.\ Nucl.\
Phys.\ \textbf{45}, S1 (2000). \vspace*{-0.5\baselineskip}

\item \label{revreinhard} R.\ Alkofer and L.\,v.\ Smekal,
Phys.\ Rept.\ {\bf 353}, 281 (2001). \vspace*{-0.5\baselineskip}

\item \label{revpieter} P.\ Maris and C.\,D.\ Roberts,
Int.\ J.\  Mod.\ Phys.\ E {\bf 12}, 297 (2003). \vspace*{-0.5\baselineskip}

\item \label{mandar} M.S.\ Bhagwat, M.A.\ Pichowsky, C.D.\ Roberts and P.C.\
Tandy, Phys. Rev. C {\bf 68}, 015203 (2003). \vspace*{-0.5\baselineskip}

\item \label{bowman2} P.O.\ Bowman, U.M.\ Heller, D.B.\ Leinweber and A.G.\ Williams,
``Modelling the quark propagator,'' hep-lat/0209129.
\vspace*{-0.5\baselineskip}

\item \label{pctlattice} P.\ Maris, A.\ Raya, C.D.\ Roberts and S.M.\ Schmidt,
``Facets of confinement and dynamical chiral symmetry breaking,''
nucl-th/0208071;
P.C.\ Tandy, Prog.\ Part.\ Nucl.\ Phys.\  {\bf 50}, 305 (2003).
\vspace*{-0.5\baselineskip}

\item \label{alkofer} C.S.\ Fischer and R.\ Alkofer,
Phys.\ Rev.\ D {\bf 67}, 094020 (2003); and references therein.
\vspace*{-0.5\baselineskip}

\item \label{latticegluon} P.O.\ Bowman, U.M.\ Heller, D.B.\ Leinweber and
A.G.\ Williams,
Phys.\ Rev.\ D {\bf 66}, 074505 (2002); and references therein.
\vspace*{-0.5\baselineskip}

\item \label{ayse} J.I.\ Skullerud, P.O.\ Bowman, A.\ {K\i z\i lers\"u},
D.B.\ Leinweber and A.G.\ Williams,
JHEP {\bf 0304}, 047 (2003). \vspace*{-0.5\baselineskip}

\item \label{detmold} A.\ Bender, W.\ Detmold, C.D.\ Roberts and A.W.\ Thomas,
Phys.\ Rev.\ C \textbf{65}, 065203 (2002). \vspace*{-0.5\baselineskip}

\item \label{mrt98} P.\ Maris, C.D.\ Roberts and P.C.\ Tandy, Phys.\
Lett.\ B\textbf{420}, 267 (1998). \vspace*{-0.5\baselineskip}

\item \label{hqlimit}
M.A.\ Ivanov, Yu.L.\ Kalinovsky and C.D.\ Roberts, Phys.\ Rev.\ D{\bf 60},
034018 (1999). \vspace*{-0.5\baselineskip}

\item \label{volkov} M.K.\ Volkov and V.L.\ Yudichev,
Phys.\ Part.\ Nucl.\  {\bf 31}, 282 (2000) [Fiz.\ Elem.\ Chast.\ Atom.\ Yadra
{\bf 31}, 576 (2000)].\vspace*{-0.5\baselineskip}

\item \label{maristandy1} P.\ Maris and P.C.\ Tandy,
Phys.\ Rev.\ C {\bf 60}, 055214 (1999). \vspace*{-0.5\baselineskip}

\item \label{piN} J.C.R~Bloch, A.~Krassnigg and C.D.\ Roberts,
``Regarding proton form factors,'' nucl-th/0306059, and references therein.

\end{enumerate}

\end{document}